\documentclass[twocolumn]{article}
\usepackage{amsmath}
\usepackage{times}
\usepackage[top=0.25in, bottom=0.5in, left=0.7in, right=0.7in]{geometry}
\usepackage{natbib}
\usepackage{graphicx}
\DeclareMathOperator{\var}{var}
\begin{document}
\title{When to Smell in Stereo}
\author{
    Sina Tootoonian\thanks{Corresponding author: \texttt{sina.tootoonian@crick.ac.uk}} \qquad  
    Andreas T. Schaefer\\
    Sensory Circuits and Neurotechnology Laboratory\\
    The Francis Crick Institute, London, UK
}
\date{}
\maketitle
\begin{abstract}
Recent works have highlighted the use of dual nostril `stereo' olfaction by a variety of animals. In this work we perform ``back of the envelope'' calculations to determine when stereo olfaction is useful compared to single nostril `mono' olfaction. We find that stereo olfaction is advantageous when there are large relative changes in the odour concentration, and when the spatial length scales of correlations in air are large, such as in the boundary layer near surfaces. In other words, stereo olfaction is useful when animals are searching surfaces for olfactory edges, such as when tracking odour trails.
  
\end{abstract}

\section{Introduction}
The bilateral symmetry of most animals has endowed many with two sets of olfactory signal detectors. For example, most insects have two olfactory appendages, their antennae, and all terrestrial mammals have two nostrils. These provide their host animals with two related, yet distinct, olfactory signals. Recent works have investigated how such `stereo' signals can influence olfactory behaviours and computations. \citep{rajan_rats_2006} showed that rats can rapidly localize odours and that performance dropped to near chance after stitching of either nostril. \citep{porter_mechanisms_2007} showed that humans performing an olfactory tracking task performed better when using both nostrils than when the same olfactory volume was presented to each nostril. \citep{khan_rats_2012} studied rats tracking a trail and found that deviations from the track increased when a nostril was stitched, while \citep{jones_mice_2018} found that the rate at which mice followed a rewarded odour trail slowed, and their deviations from the trail increased, when a nostril was occluded. Similarly, \citep{jayakumar_mice_2026} observed that mice at the edges of a trail they were tracking would quickly reorient towards it, but much less readily if their trail-side nostril was blocked. \citep{catania_stereo_2013} studied location of food sources in a gradient by blind moles and found that they too benefit from binaral olfaction, particularly when closer to the source, when gradients are strongest.




These experimental findings highlight the importance of binaral signals for performing important olfactory tasks. Theoretical and computational works have begun to investigate the conditions under which binaral signals are beneficial, and the computations to be performed on them. \citep{boie_information-theoretic_2018} used information theory to establish the advantage of distributing a fixed coding budget among two spatially separated samples, rather than to increasing the resolution of a single sample, for encoding location within an odour plume. In that work, the two samples were coded independently, so in a follow-up study, \citep{victor_information-theoretic_2022} investigated the joint coding of the two samples. They found that optimal encoding depended on the flow conditions: in turbulent flows, optimal codes encoded the maximum of the two samples, while in smoother, boundary layer flows, optimal codes encoded the difference of the two samples.

In this work, we present ``back of the envelope'' calculations that suggest the conditions when stereo olfaction is advantageous. We were inspired by the recent experimental study of \citep{asumbisa_stereo_2025} suggesting that stereo olfaction is crucial for head-direction coding in mice. We  compute the signal to noise ratios (SNRs) for both the summed and differential signals, and compare them to determine when stereo olfaction is more informative.
\section{Results}
\begin{figure}
  \centering
  \includegraphics[width=0.5\textwidth]{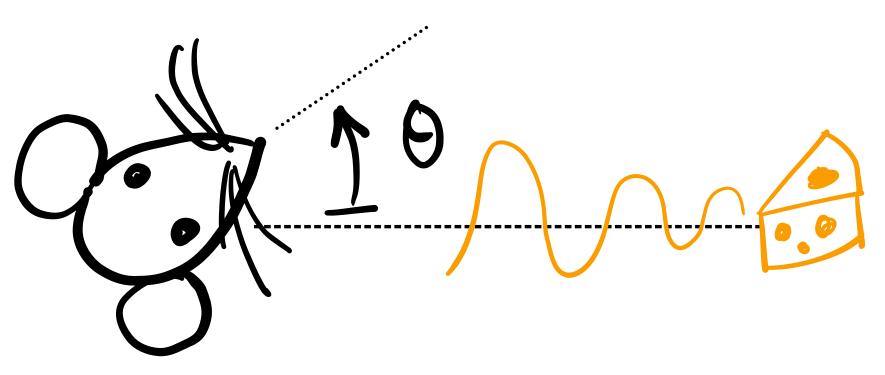}
  \caption{A mouse facing an odour source. The angle the mouse's head makes relative to the odour source is $\theta$, zero when facing directly upstream and increasing counterclockwise.}
  \label{fig:mouse_cheese}
  \end{figure}

We consider a mouse facing upstream towards an odour source (Fig. \ref{fig:mouse_cheese}). Let $\theta$ be the angle the mouse's head makes relative to the odour source, zero when facing directly upstream and increasing counterclockwise.
 The odour concentration will vary as the mouse rotates its head. Let the corresponding mean concentration be $\mu(\theta)$. We assume odour packet arrivals are memoryless at this mean rate. We therefore model the odour signal present in a single sniff as a Poisson process, \begin{align} r(\theta) \sim \text{Poisson} (\mu(\theta)).\end{align}
An olfactory system using a single nostril would only have the single rate signal above to work with. We can therefore define a mono-nostril, per-sniff signal-to-noise ratio using the mean and standard deviation of this signal, \begin{align}\label{mono_snr}\text{SNR}_\text{mono} = {\mu(\theta) \over \sqrt{\mu(\theta)}} = \sqrt{\mu(\theta)}.\end{align}
\subsection{Stereo olfaction}
For the comparison to stereo olfaction, we will further assume that the rate of odour-packet arrivals is fast enough that we can approximate the Poisson process with a Gaussian. Thus, we can describe the odour signal as the sum of a mean concentration and a Gaussian noise term, $$ r(\theta) = \mu(\theta) + \eta, \quad \eta \sim \mathcal{N}(0, \mu(\theta)).$$

We can then consider stereo olfaction as two such signals, at angles of $\pm \Delta \theta/2$ relative to the head direction. The signal at the first nostril is \begin{align*} r_1(\theta) = \mu\left(\theta - {\Delta \theta \over 2}\right) + \eta_1, \quad \eta_1 \sim \mathcal{N}\left(0, \mu\left(\theta - {\Delta \theta \over 2}\right)\right).\end{align*}

If we assume small inter-nostril distance relative to the length-scale over which concentrations change, we can approximate the signal at the first nostril as $$ r_1(\theta) \approx \mu(\theta) - \mu'(\theta) {\Delta \theta \over 2} + \eta_1, \quad \eta_1 \sim \mathcal{N}(0, \mu(\theta)).$$
The signal at the second is similar but at the angle of the second nostril, so $$ r_2(\theta) \approx \mu(\theta) + \mu'(\theta) {\Delta \theta \over 2} + \eta_2, \quad \eta_2 \sim \mathcal{N}(0, \mu(\theta)).$$

Crucially, the noise at the second nostril is correlated with the noise at the first by the length scales of the odour plume at the nose. Letting this correlation be $\rho$, we have $$ \eta_2 = \rho \eta_1 + \sqrt{1 - \rho^2} \eta, \quad \eta \sim \mathcal{N}(0, \mu(\theta)),$$ where $\eta$ is the noise component at the second nostril that’s uncorrelated with that at the first.

\subsection{The stereo difference}
How should an animal combine the signals from the two nostrils? Averaging them would reproduce the mono-nostril signal but with reduced variance. Because of the correlated noise at the two nostrils, this variance reduction would be small.

The natural operation is to \emph{subtract} the two signals. This gives, \begin{align}\label{stereo} \Delta r(\theta) = r_2 (\theta) - r_1(\theta) = \mu'(\theta)\Delta \theta + \xi,\end{align} where the noise term is $$ \xi = -(1 -\rho) \eta_1 + \sqrt{1 - \rho^2} \eta.$$ The two noise components $\eta$ and $\eta_1$ have mean zero, and are uncorrelated so their variances add, giving $$ \text{var}(\xi) = (1 - \rho)^2 \var(\eta_1) + (1 - \rho^2)\var(\eta) = 2(1 - \rho) \mu(\theta).$$

We can then compute an SNR for stereo olfaction by comparing the mean and standard deviation of the difference signal in Eqn. \ref{stereo}, \begin{align}\label{stereo_snr}\text{SNR}_\text{stereo} = {|\mu'(\theta)| \over \sqrt{\mu(\theta)}}{\Delta \theta \over \sqrt{2 ( 1 - \rho)}}.\end{align}

\subsection{When to turn on the stereo} We can determine the relative advantage of stereo to mono olfaction as the ratio of the SNRs, \begin{align} \label{snr_ratio}{\text{SNR}_\text{stereo} \over \text{SNR}_\text{mono}} = {|\mu'(\theta)| \over \mu(\theta)}{\Delta \theta \over \sqrt{2 ( 1 - \rho)}}.\end{align} This expression combines two terms. The first says that stereo olfaction is beneficial when the relative change in the mean concentration is large. This is not surprising since we are effectively comparing a difference of two signals to their common component.

The second term in the ratio is determined by two factors. It \emph{increases} as the inter-nostril spacing $\Delta \theta$ grows and correspondingly larger concentrations changes can be observed. It \emph{decreases} as the inter-nostril correlation $\rho$ shrinks and subtraction becomes less effective in noise removal.

These two factors are related, since increasing the distance between the nostrils will locate their signals farther apart in the plume, reducing their correlations. To see this, note first that the correlation function is symmetric, so $\rho(\Delta \theta) = \rho(-\Delta \theta)$. If we assume that correlations are twice differentiable in $\Delta \theta$, then this symmetry implies that  $$ \rho(\Delta \theta) \approx 1 - {\Delta \theta^2 \over 2 L^2},$$  where $L$ is the correlation length-scale of the plume. Our SNR ratio in Eqn. \ref{snr_ratio} then simplifies to \begin{align}
  \label{snr2}{\text{SNR}_\text{stereo} \over \text{SNR}_\text{mono}} \approx {|\mu'(\theta)| \over \mu(\theta)} L.
  \end{align}

\section{Discussion}
The expression in Eqn. \ref{snr2} suggests that stereo olfaction is advantageous when two conditions are met. First, the relative change in mean odour concentrations should be large, i.e. at odour edges. Second, the length scales of odour correlations should be large, so that subtracting the signal from the two nostrils can largely remove their noise component. One prominent location of such large length scales is the laminar boundary-layer flow near surfaces. In short, stereo olfaction is most beneficial when searching surfaces for olfactory edges. 

We have made numerous simplifying assumptions to arrive at the expression in Eqn. \ref{snr2}, and the true expression for real plumes is likely to be different. Nevertheless, we expect the qualitative relationship in that expression to hold: stereo signals will be advantageous when relative concentration changes rapidly, and when plume length scales are large enough that subtraction yields noise-reduction.

The only comparative computation we considered was subtraction. Although mathematically simple, subtraction of two timeseries would be difficult to neurally implement as it would require precise temporal alignment of excitatory and inhibitory signals, one of them contralateral and therefore delayed. We avoided this problem by appealing to the temporal discretization imposed by the sniff and assuming that the quantities being compared are not moment-by-moment concentration fluctuations at each nostril, but the respective aggregate excitations caused by the odour contained in the inhaled volume per sniff. The inter-sniff interval may in turn provide enough time for the signal from one nostril to be inverted and subtracted from that from the other. 

One way to avoid the complications of interacting excitatory and inhibitory signals is to avoid inhibition entirely. For example, although the task of \citep{rajan_rats_2006} is consistent with a differential comparison of the binaral odour signal, it could in principle be performed using the total intensity signal from two asymmetric nostrils (Dima Rinberg, personal communication). However, the follow up study \citep{parthasarathy_laterality_2013}  provided evidence that the nasal signals were indeed symmetric, implying a differential computation. 

Interestingly, the subliminal binaral sensitivity \citep{wu_humans_2020} found in humans evaluating self-motion in an optic flow task was not to the difference of the nasal signals, but to their ratio. However, such ratios can be computed as the difference of logarithmically transformed signals. Consistent with this transformation, the authors found that the subjects' perceived odour intensity scaled logarithmically with concentration.

When comparing mono and stereo olfaction we've used the corresponding signal-to-noise ratios. However, the nature of the signal in the two cases is different, being the mean concentration in the first case, and change in mean concentration in second. Is such a comparison meaningful? We view this through the lens of signal detection. The mono and stereo olfaction signals reflect the components of the odour signal at low and high spatial frequencies. The SNR comparison is then a measure of when the high-frequency component of this signal is more informative than the low-frequency component.

A concrete example of this is when considering strategies for performing the tracking task in \cite{porter_mechanisms_2007}. In the first, the tracker would sum the binaral signal to estimate $\mu(\theta)$ to estimate odour presence. In the second, the tracker would use the difference of the binaral signal to estimate $\mu'(\theta)$. Our calculation predicts that the stereo signal would be advantageous on a calm day when length scales are large, while the mono signal would be advantageous on a windy day, when length scales shrink. However, the optimal Bayesian choice is not to use one signal or the other, but to perform a precision-weighted combination of the two signals.

Our calculation highlights the relation between the inter-nostril spacing and the length-scales of the plume. It suggests that stereo olfaction would be more important in surface-bound animals like mice, who benefit from the large length scales at surfaces. In contrast, for animals who typically sample the bulk air volume above surfaces, stereo olfaction is likely less beneficial, since flow in such settings is turbulent and has smaller correlation length scales. This effect can be mitigated if the nostrils are closer together, for example in flying insects, some species of which have been shown to use stereo olfactory signals (see e.g. \cite{kadakia_odour_2022}). 

\section*{Acknowledgements}
We are grateful to Jonathan Victor, Dima Rinberg, and members of the Schaefer lab for their feedback on this manuscript.  This work was supported by the Francis Crick Institute, which receives its core funding from Cancer Research UK (CC2036 to ATS), the UK Medical Research Council (CC2036 to ATS), and the Wellcome Trust (CC2036 to ATS), and by the National Science Foundation / Canadian Institutes of Health Research / German Research Foundation / Fonds de Recherche du Qu\'ebec / UK Research and Innovation–Medical Research Council Next Generation Networks for Neuroscience Program (Award No. 2014217 to ATS).
\bibliographystyle{plainnat}
\bibliography{ms_clean}
\end{document}